\documentclass[12pt]{article}
\usepackage{epsfig,amsmath,amssymb,array,dcolumn,subfigure,rotating}
\usepackage[a4paper]{geometry}

\def\ul#1#2{\textstyle{\frac#1#2}}
\newcommand {\vct}[1] {\mathbf {#1}}
\begin{document}

\title{Electrostatic Contribution to the Persistence Length of a Semiflexible Dipolar Chain}

\author{Rudi Podgornik \\[3mm] 
Department of Physics, University of Ljubljana,\\
Jadranska 19, 1000 Ljubljana, Slovenia and \\
Department of Theoretical Physics, J. Stefan Institute,\\
Jamova 39, SI-1000 Ljubljana, Slovenia \\
and\\
Laboratory of Physical and Structural
Biology\\
NICHD, Bld. 9 Rm. 1E116 \\
National Institutes of Health, Bethesda, MD
20892-0924 }

\maketitle

\begin{abstract}
We investigate the electrostatic contribution to the persistence length of a semiflexible polymer 
chain whose segments interact via a screened Debye-H\" uckel dipolar interaction
potential. We derive the expressions  for the renormalized persistence
length on the level of a $1/D$-expansion method already successfully used in
other contexts of polyelectrolye physics. We investigate different limiting forms
of the renormalized persistence length of the dipolar chain and show that in
general it depends less strongly on the screening length than in the context
of a monopolar chain. We show that for a dipolar chain the electrostatic persistence
length in the same regime of the parameter phase space as the original
Odijk-Skolnick-Fixman (OSF ) form for a monopolar chain  depends
logarithmically  on the screening length rather than quadratically. This can be
understood solely on the basis of a swifter decay of the dipolar interactions
with separation compared to the monopolar  electrostatic interactions. We comment also on 
the general contribution of higher multipoles to the electrostatic renormalization  of the
bending rigidity.
\end{abstract}

\section{Introduction}

Semiflexible polyelectrolytes are ubiquitous in biological context ranging from biopolymers such as 
DNA, filamentous $F$ actin, microtubules, and then all the way to molecular aggregates such as micelles 
or  even whole organisms as in bacterial $fd$ viruses or the tobacco mosaic virus. In all these cases we 
are dealing with objects that at an appropriate scale behave as Euler-Kirchhoffian
elastic filaments. The  most important mechanical characteristics in these
molecular systems is the persistence length stemming from 
the bending rigidity of the polymer that can exhibit enormous variation in magnitude bracketted by 
nm on the lower end and mm on the upper end.

The bending rigidity and thus the persistence length is a consequence of short
range atomic and molecular interactions and is itself a mesoscopic property
\cite{yamakawa}. For charged semiflexible polymers the long range nature of the
electrostatic interactions modifies the value of the persistence length as is
well known from the seminal work of Odijk, Skolnick and Fixman (OSF) \cite{OSF}.
Recent detailed critical assessment of the OSF conjecture \cite{Everaers,
Ullner, Shklovski} confirmed the universality of the dependence of the
persistence length  on the parameters of the electrostatic interaction. It appears
that the OSF behavior characterized by the inverse dependence of the persistence
length on the ionic strength of the bathing medium is quite robust. One of the
main ingredients of the OSF ideology is a complete lack of any atomic 
or molecular specificity in the interaction between charge segments
along the polyelectrolyte. By assumption the mobile charges in solution and the
fixed charges on the polyelectrolyte backbone interact only via generic
(screened) Coulomb interaction.
Though this assumption seems to be reasonable it would be appropriate to
investigate what kind of behavior of persistence length is conferred by the
effects of strong specificity as in {\sl e.g.} the case of specific ion
adsorption. Recent experiments also suggest that reevaluation of the basic
assumptions in our understanding of counterion - backbone interactions in
polyelectrolytes would be highly desirable \cite{hoagland}.

In specific ion adsorption mobile charges from the bathing solution and fixed
charges along the polyelectrolyte can combine, leading to the emergence of higher
multipoles along the polyelectrolyte chain, the first one being a dipole
stemming from the association of a negative fixed charge and a specifically
adsorbed mobile positive charge from the bathing solution.  The same situation though for different
reasons, specifically due to the low dielectric constant envoronment, could be obtained also in
the context of ionomers \cite{ionomers}. Similar consideration could apply also to
electrorheological or ferrofluids in external fields.

Muthukumar \cite{Muthu} was the first to realize the importance and the extent of
modifications wrought by the first higher multipole {\sl i.e.} the dipole, on the
behavior of a {\sl flexible} polyelectrolyte described on the level of the
Edwards Hamiltonian. He discovered the formation of
localized aggregated structures along the chain that dominate the statistical
behavior of the flexible polyelectrolyte chain. They are characterized by a
different scaling of the size of the chain with respect to its length and depend
continuously on the parameters of the dipolar interaction potential.

Chain flexibility as formalized by the Edwards Hamiltonian \cite{Muthu1} is
essential for emergence of this type of aggregated structures. Now assume a 
dipolar {\sl semiflexible} chain, described  as an Euler-Kirchhoffian filament
with dipolar intersegment interactions.  On top of that assume that the length of
the chain is the largest length in the problem. This constitutes a
complementary limit to be contrasted with the Muthukumar calculation.
The analysis provided by Muthukumar and the one detailed below should thus
brackett the behavior of any real polyelectrolyte chain with dipolar charges
along its length. 

The problem with dipolar interactions is that they depend on the local
orientation of the polyelectrolyte segments. This complicates the evaluation
of the partition function of the semiflexible chain in an essentail manner. 
In order to make this complicated problem tractable, we will invoke 
the $1/D$-expansion method \cite{Polyakov}  that has already been successfully
applied to the problem of persistence length renormalization of a semiflexible
polyelectrolyte  chain \cite{Hansen} with monopolar charges. This will allow
us to explicitly evaluate the electrostatic contribution to the persistence
length of a dipolar chain. As a point of departure we will assume
that the dipolar charges along the chain interact {\sl via} a screened Coulombic
interaction potential, the range of which is characterized by the Debye screening
length \cite{Barrat}. We will derive the complete dependence of the
electrostatic persistence length on the parameters of the dipolar interaction
potential and show that it leads to a different behavior regarding its dependence
on the ionic strength of the bathing solution then the standard
OSF result. We find basically two different regimes for the bahavior of the
electrostatic persistence length of a dipolar chain
\begin{itemize}
\item for small but non-vanishing values of the screening length $\lambda_{D}$
the electrostatic contribution to the persistence length behaves as
$$L_{P}^{(R)} = L_{P}^{(0)} + const. \log{\lambda_{D}}$$
where $L_{P}^{(0)}$ is the bare value of the persistence length. This
expression would superseed the OSF $\lambda_{D}^{2}$ result valid for
monopolar interactions. Obviously the dipolar electrostatic renormalization of
the persistence length depends more gently on the screening length. 
\item for large values of the screening length $\lambda_{D}$ the electrostatic
contribution to the persistence length behaves as $$
L_{P}^{(R)} = const. {\lambda_{D}}^{\beta},$$where $\beta$ equals either $3/4$ or 
$3$, depending on the strength of the dipolar interaction. This result should be
compared with various variational estimates of a sub-OSF behavior for a monopolar
chain that lead to $\beta$ in the vicinity of $1$.
\end{itemize}
These two regimes stemm from two different limiting forms of a single equation
giving the renormallized value of the persistence length as a function of the
parameters describing the dipolar interactins along the chain. We also present
complete numerical solutions of this equation in various regions of the
parameter space.

The outline of the paper is as follows. In Section 1 we first rewrite the
Hamiltonian of a semiflexible polymer chain
with screened dipolar interactions, assumed to be composed of the
Euler-Kirchhoff elastic energy and the interaction energy, in a form that allows
for a straightforward application of the $1/D$-expansion ansatz. We assume that
the dipoles of the polymer segments are oriented along the local tangent vectors
of the chain. This effective Hamiltonian captures the elasticity of the chain,
the inextensibility of the chain and the fact that electrostatic interactions
depend on the position of the interacting segments as well as their orientation.
Section 2
introduces all the important approximations in order to make the evaluation of
the partition function of the chain tractable. A diagonalization ansatz is
introduced for the orientational part of the Hamiltonian, a global
inextensibility constraint is substituted for the local one and a saddle point
evaluation is introduced for all the auxiliary fields that do not enter the
Hamiltonian on a quadratic level. In section 3 the saddle point equations are
solved explicitly by expanding all the Fourier components of different
auxiliary fields to the fourth order. This is consistent with the semiflexible
ansatz for the configuraional part of the Hamiltonian of the chain. Section 4
introduces explicit equations for the bending rigidity renormalization that
follow from the saddle point equations of the previous section. These
equations are solved numerically in the last section. Different limiting forms
of the numeric solutions for limited regions of the parameter phase space are
derived also analytically. We take a critical look at the results derived in
this work and comment on the implied limitations of their validity and
their connection  with previous work on the electrostatic renormalization of
the rigidity of semiflexible polymers. 

\section{The effective Hamiltonian}

We investigate a semiflexible polymer chain with dipolar charges along the
chain. The dipoles can be either structural or they can be a consequence of the
specific adsorption of {\sl e.g.} positive mobile charges from the bathing
solution onto the fixed negative charges on the polyelectrolyte backbone. The
total interaction energy of the chain is given by
\begin{equation}
\frac12 \int_{0}^{L}\!\!\!\int_{0}^{L} V({\bf r}(s), {\bf r}(s')) ds ds'
\end{equation}
where 
\begin{eqnarray}
V({\bf r}(s), {\bf r}(s')) &=& \frac{e^{-\kappa \vert {\bf r}(s) - {\bf
r}(s')\vert}}{4 \pi
\epsilon\epsilon_{0} \vert {\bf r}(s) - {\bf r}(s')\vert^{5}} \left( \vert {\bf r}(s) - {\bf r}(s')\vert^{2} 
(1 + \kappa \vert {\bf r}(s) - {\bf r}(s')\vert) \vct p(s)\cdot\vct p(s') -
\right.
\nonumber\\
& & \left. - (3 + 3 \kappa \vert {\bf r}(s) - {\bf r}(s')\vert + (\kappa \vert
{\bf r}(s)- {\bf r}(s')\vert)^{2}) (\vct p(s)\cdot ({\bf r}(s) - {\bf r}(s')) )
(\vct p(s')\cdot ({\bf r}(s) - {\bf r}(s')))\right),
\end{eqnarray}
with $\vert {\bf r}(s) - {\bf r}(s')\vert$ the separation between two segments
located at ${\bf r}(s)$ and ${\bf r}(s')$ along the chain,  $s$ is the
arclength of the chain, and  $\vct p(s)$ and $\vct p(s')$ are dipoles per
unit length located at $s$ and $s'$.  The above form of the screened
dipolar interaction follows straightforwardly from the second order multipole
expansion of the screened Coulomb kernel \cite{Schwinger} and reduces to the usual form of the
dipolar interaction in the lim it of no screening. On the Debye - H\" uckel level  $\kappa =
\lambda_{D}^{-1}$ where $\lambda_{D}$ is the Debye screening length. We
furthermore assume that the dipole per unit length along the chain is given by
\begin{equation}
{\bf p}(s) = p_0 \partial_{s} {\bf r}(s)
\end{equation}
where $p_0$ is the strength of the dipolar moment per unit length of the
segment and $\partial_s {\bf r}(s) = \dot{\bf r}(s)$ is the unit tangent vector.
It is thus assumed that the dipoles point along the
chain, and  that the component of the dipoles perpendicular to the local axis of
the chain is averaged out to zero. Or model is thus "perpendicular" to ther case investigated in the
context of protein folding \cite{orland}. The interactions along the chain are
therefore described with a generic form of segment - segment  interactions that
can be cast into the form
\begin{eqnarray}
V({\bf r}(s), {\bf r}(s')) &=& V_R({\bf r}(s) - {\bf r}(s'))  \partial_s {\bf r}(s)  \partial_{s'}
{\bf r}(s') -  \nonumber\\ 
& & - V_P({\bf r}(s) - {\bf r}(s')) \left( \partial_s {\bf r}(s) \cdot ({\bf r}(s) - {\bf r}(s'))
\right) \left ( \partial_{s'} {\bf r }(s') \cdot ({\bf r}(s) - {\bf r}(s')) \right), \nonumber\\
~
\label{eq.1.0}
\end{eqnarray}
with
\begin{eqnarray}
V_R({\bf r}(s) - {\bf r}(s')) &=& v_0 \frac{e^{- \kappa \vert {\bf r}(s) - {\bf r}(s')\vert}}{\vert{\bf r}(s) - {\bf r}(s')\vert^3} \left( 1 + \kappa \vert{\bf r}(s) - {\bf r}(s')\vert \right), \nonumber\\
V_P({\bf r}(s) - {\bf r}(s')) &=&  v_0 \frac{e^{- \kappa \vert {\bf r}(s) - {\bf r}(s')\vert}}{\vert{\bf r}(s) - {\bf r}(s')\vert^5} \left( 3 + 3 \kappa \vert{\bf r}(s) - {\bf r}(s')\vert + \kappa^2 \vert{\bf r}(s) - {\bf r}(s')\vert^2\right),
\end{eqnarray}
Obiosuly the interaction Eq. \ref{eq.1.0} depends on the positions of the
interacting segments as well as on their orientation. This is the fundamental
difference between monopolar and dipolar interactions. We have defined
\begin{equation}
v_0 = \frac{p_0^2}{4\pi \epsilon \epsilon_0} = k_B T~~\ell_B \left( \frac{p_0^2}{e_0^2} \right),
\label{defv0}
\end{equation}
where $\ell_B$ is the Bjerrum length \cite{Barrat}.  The units of $v_0$ are
energy times length and are thus the same as the units  of bending modulus of a
semiflexible polymer chain. 

The generic form of the interaction Eq. \ref{eq.1.0} together with the
Euler-Kirchhoffian elastic conformational part of the mesoscopic free energy
\cite{Kamien} can now be used
to investigate the statistical behavior of a semiflexible dipolar chain. The mesoscopic
Hamiltonian of the chain  can  be written canonically as
\begin{equation}
{\cal H}[{\bf r}(s)] = \frac12 K_C \int_{0}^{L} ds \left( \partial^2_s {\bf r}(s) \right)^2 +
\frac12 \int_{0}^{L}\!\!\!\int_{0}^{L} ds ds'
V({\bf r}(s), {\bf r}(s')).
\end{equation}
where the local curvature is
\begin{equation}
\partial^2_s {\bf r}(s) = \frac{\partial^2 {\bf r}(s)}{\partial s^2}.
\end{equation}
The bending rigidity $K_{C}$ is connected with the bare persistence length
{\sl via} $K_{C} = k_{B}T~L_{P}^{{(0)}}$. Since in what follows the chain is
assumed to be inextensible an additional constraint of $ 
\partial_s {\bf r}(s) 
\partial_s {\bf r}(s) = 1$ should be taken into account. The partition function
thus assumes the form
\begin{equation}
\Xi = \int {\cal D}[{\bf r}(s)] \Pi_s \delta^3\left( (\partial_s {\bf r}(s))^2 - 1 \right) ~\exp{-\beta {\cal H}[{\bf r}(s)]}.
\end{equation}
There are two difficult problems connected with this partition function: first of all
we have the constraint of inextensibility
that has to be enforced locally for every conformation of the chain, and on top of this
there is the non-locality of the
segment-segment interaction potential that couples different segments along the
chain and depends also on their orientation. We will address  both problems
systematically by applying the method of Lagrange multipliers \cite{Hansen}. 

To this effect we will introduce two additional terms into the  interaction Hamiltonian. The inextensibility
constraint is dealt with {\sl via} the Lagrange multiplier $\lambda(s)$, leading
to an additional term in the Hamiltonian of the form
\begin{equation}
\delta {\cal H}_1 = \frac{1}{2} \int ds \lambda(s) \left( (\partial_s {\bf r}(s))^2 - 1
\right),
\end{equation}
together with an additional trace over $\lambda(s)$ in the expression for the
partition function. In what follows we will assume that the inextensiblity
constraint can be implemented globally instead of locally \cite{Edwards}. This
automatically means that $ \lambda(s) =  \lambda$. The non-locality of the
interaction potential is more difficult to  deal with then in the case of a chain
without orientationally dependent interactions. In order to pave the way for an
approximate evaluation of the partition function we introduce two new Lagrange
multipliers in the form of two tensorial auxiliary fields. First of all we define
\begin{equation}
{\cal B}_{ik}(s,s') = \left( {\bf r}_i(s) -  {\bf r}_i(s')\right)  \left( {\bf r}_k(s) -  {\bf r}_k(s')\right)  
\end{equation}
and then
\begin{equation}
{\cal T}_{ik}(s,s') = \partial_s {\bf r}_i(s)\partial_s {\bf r}_k(s).
\end{equation}
We notice immediately these two fundamental identities 
\begin{equation}
{\rm Tr}~ {\cal B}_{ik}(s,s') =  \left( {\bf r}(s) -  {\bf r}(s')\right)^2 \quad {\rm and } \quad {\rm Tr}~{\cal T}_{ik}(s,s') = \partial_s {\bf r}(s)\cdot\partial_s {\bf r}_(s).
\end{equation}
We thus conclude that ${\rm Tr}~ {\cal B}_{ik}(s,s')$ is equivalent to ${\cal
B}(s,s')$ introduced in the context of orientationaly independent intrachain
interactions \cite{Hansen}. Introducing furthermore the following two additional
coupling terms
\begin{eqnarray}
\delta{\cal H}_2 &=& \frac{1}{2} \int\!\!\int ds ds' g_{ik}(s,s') \left(  
\left( {\bf r}_i(s) -  {\bf r}_i(s')\right)  \left( {\bf r}_k(s) -  {\bf
r}_k(s')\right) - {\cal B}_{ik}(s,s') \right) \nonumber\\
\delta{\cal H}_3 &=& \frac{1}{2}  \int\!\!\int ds ds' p_{ik}(s,s') \left(  
\partial_s {\bf r}_i(s)\partial_s  {\bf r}_k(s) - {\cal T}_{ik}(s,s') \right). 
\end{eqnarray}
we can write the partition function in the following form
\begin{equation}
\Xi = \int {\cal D}[{\bf r}(s)] {\cal D}[\lambda(s)] {\cal D}[g_{ik}(s)] 
{\cal D}[{\cal B}_{ik}(s)]  {\cal D}[p_{ik}(s)] {\cal D}[{\cal T}_{ik}(s)]  
~\exp{-\beta \overline{\cal H}[{\bf r}(s)]},
\label{parti}
\end{equation}
where the effective Hamiltonian $\overline{\cal H}[{\bf r}(s)] $ is given by
\begin{equation}
\overline{\cal H}[{\bf r}(s)]  = {\cal H}[{\bf r}(s)]+ \delta {\cal H}_1[{\bf r}(s)] + 
\delta{\cal H}_2[{\bf r}(s)] + \delta{\cal H}_3[{\bf r}(s)].
\end{equation}
The new auxiliary fields $g_{ik}(s,s')$ and $p_{ik}(s,s')$ thus act as
tensorial Lagrange multipliers setting the constraints $\left( {\bf r}_i(s) -
 {\bf r}_i(s')\right)  \left( {\bf r}_k(s) -  {\bf
r}_k(s')\right) - {\cal B}_{ik}(s,s') = 0$ and $\partial_s {\bf r}_i(s)\partial_s  {\bf r}_k(s) - {\cal
T}_{ik}(s,s') = 0$. In the new variable space the segment-segment interaction
potential can be obtained in a more transparent form as
\begin{eqnarray}
V({\bf r}(s), {\bf r}(s')) &=& V_R({\rm Tr}~ {\cal B}_{ik}(s,s')) {\rm Tr}~{\cal T}_{ik}(s,s') - \nonumber\\
& & - V_P({\rm Tr}~ {\cal B}_{ik}(s,s')) {\cal B}_{ik}(s,s') {\cal T}_{ik}(s,s') = V\left( {\cal B}_{ik}(s,s'), {\cal T}_{ik}(s,s') \right).
\end{eqnarray}
In order to arrive at a more transparetn form of the partition function we
furthermore introduce the following two new variables
\begin{eqnarray}
\lambda^{(c)}_{ik}(s,s') &=& \lambda ~\delta_{ik} \delta(s -s') + p_{ik}(s,s')
\nonumber\\
g^{(c)}_{ik}(s,s') &=& \frac{1}{2} \delta(s -s') \int ds" \left( g_{ik}(s,s") + g_{ik}(s",s')\right) - g_{ik}(s,s').
\end{eqnarray}
With these new definitions the effective Hamiltonian can be finally reduced to
this fairly complicated form
\begin{eqnarray}
\overline{\cal H}[{\bf r}(s)] &=& \frac{1}{2} K_C \int ds (\partial^2_s{\bf r}(s))^2 - \ul12 \lambda \int ds +\nonumber\\
& & + \frac{1}{2} \int\int ds ds' \lambda^{(c)}_{ik}(s,s')  \partial_s {\bf r}_i(s)  \partial_s {\bf r}_k(s) + \frac{1}{2} \int\int ds ds' 2 g^{(c)}_{ik}(s,s') {\bf r}_i(s){\bf r}_k(s) - \nonumber\\
& & - \frac{1}{2}  \int\int ds ds' g_{ik}(s,s'){\cal B}_{ik}(s,s') - \frac{1}{2} \int\int ds ds' p_{ik}(s,s') {\cal T}_{ik}(s,s') + \nonumber\\
& & + \frac{1}{2} \int\int ds ds' V\left( {\cal B}_{ik}(s,s'), {\cal T}_{ik}(s,s') \right).
\label{first-ham}
\end{eqnarray}
The above form of the Hamiltonian together with Eq. \ref{parti} represents at this stage
an exact expression for the partition function. This is the starting point of
different approximations introduced below. 

Though the above form of the effective Hamiltonian looks prohibitively
complicated it can in fact be reduced to analytical quadratures provided one
devises a powerfull enough approximation scheme. It was shown in recent work
\cite{Hansen, Podgornik, Hansen2}  that the $1/D$-expansion method, where $D$
here and below is the dimensionality of the embedding space,  can be fruitfully
applied to polymer problems of the above type and we will use our experience
gained in the context of monopolar interactions \cite{Hansen} to tackle also the
more complicated case of multipolar interactions as implied by the Hamiltonian
Eq. \ref{first-ham}.

\section{The {\bf\sl ansatz}}

Our rationale for writing the effective Hamiltonian in the form Eq. \ref{first-ham} is that it will be shown
to be amenable to straightforward approximations leading to a closed form
evaluation of the partition function. Similar
methods have been already used successfully in the context of monopolar interactions.
First of all we will introduce a {\sl diagonalization ansatz} of  the form
\begin{eqnarray}
g_{ik}(s,s') &=& g(s,s') \delta_{ik} \nonumber\\
p_{ik}(s,s') &=& p(s,s') \delta_{ik}.
\end{eqnarray}
Though this {\sl ansatz} is not absolutely necessary in order to make 
progress in the evaluation of the partition function Eq. \ref{parti}, it
certainly makes the whole calculation manageable and controllable, reducing its
overall complexity to a bare minimum. The above {\sl ansatz} clearly implies
also
\begin{eqnarray}
g^{(c)}_{ik}(s,s') &=& g^{(c)}(s,s') \delta_{ik} \nonumber\\
\lambda^{(c)}_{ik}(s,s') &=& \lambda^{(c)}(s,s') \delta_{ik}.
\end{eqnarray}
In order to be consistent we must also introduce the same type of {\sl
diagonalization ansatz} for the fields  ${\cal B}_{ik}(s,s')$ and  ${\cal
T}_{ik}(s,s')$, which are conjugate to the fields $g_{ik}(s,s')$ and
$p_{ik}(s,s')$. We formulate this as 
\begin{eqnarray}
{\cal B}_{ik}(s,s') &=& \ul1D {\cal B} (s,s') \delta_{ik} \nonumber\\
{\cal T}_{ik}(s,s') &=&\ul1D  {\cal T}(s,s') \delta_{ik}.
\end{eqnarray}
In this way we can write ${\rm Tr}~ {\cal B}_{ik}(s,s') = {\cal B}(s,s')$ and
${\rm Tr}~ {\cal T}_{ik}(s,s') = {\cal T}(s,s')$. Since we are not concerned
here with explicit analysis of finite size effects we can assume overall that
the $(s,s')$ dependencies actually reduce to $(s-s')$. Similarly to the case
of orientationally independent intrachain interactions we see that the
partition function is now quadratic in ${\bf r}(s)$ and its first and second 
derivatives. We can thus trace over these harmonic degrees of freedom
\cite{Hansen}. Expanding
the whole effective Hamiltonian  around a reference configuration ${\bf r}_0(s)$
and introducing the Fourier components of different fields  in the standard way
we obtain after tracing out the harmonic degrees of freedom the following form
of the effective Hamiltonian
\begin{eqnarray}
\overline{\cal H}' &=& \overline{\cal H}[{\bf r}_0(s)] + \textstyle{\frac{D k_B T~}{2}} \sum_{Q} \log{{\cal G}(Q)} - \frac{1}{2} \lambda \int ds +\nonumber\\
& & - \frac{1}{2}  \int\int ds ds' g(s,s'){\cal B}(s,s') - \frac{1}{2} \int\int ds ds' p(s,s') {\cal T}(s,s') + \nonumber\\
& & + \frac{1}{2} \int\int ds ds' V\left( {\cal B}(s,s'), {\cal T}(s,s') \right),
\label{eq.1.1}
\end{eqnarray}
where  the {\sl self energy} function ${\cal G}(Q)$ is given by
\begin{equation}
{\cal G}(Q) = \left(  K_C Q^4 +  \lambda^{(c)}(Q) Q^2 + 2 g^{(c)}(Q)\right).
\label{eq.1.2}
\end{equation}
Here $\lambda^{(c)}(Q)$ and  $g^{(c)}(Q)$ are of course the Fourier components
of the fields $g^{(c)}(s - s')$ and $\lambda^{(c)}(s - s')$. The Hamiltonian of
the reference configuration  ${\bf r}_0(s)$ can be derived as
\begin{eqnarray}
\overline{\cal H}[{\bf r}_0(s)] &=& \frac{1}{2} K_C \int ds (\partial^2_s{\bf r}_0(s))^2 
+ \frac{1}{2} \int\!\!\!\int ds ds' \lambda^{(c)}(s-s')  \left( \partial_s {\bf
r}_0(s) \cdot \partial_s {\bf r}_0(s)\right) + \nonumber\\
& & + \frac{1}{2} \int\!\!\!\int ds ds' 2 g^{(c)}(s-s') \left( {\bf
r}_0(s)\cdot{\bf r}_0(s) \right).
\end{eqnarray}
All along we disregard the finite (chain) size effects and assume
that the chain is homogeneous and  $s,s'$ dependence is actually equivalent to 
$\vert s - s'\vert$ dependence. This assumption furthermore implies that the
length of the chain is the largest length in the problem. Thus with this line
of reasoning
\begin{eqnarray}
g^{(c)}(Q) &=& \int du \cos{Qu }~ g^{(c)}(u) = g(Q) - g(Q=0)\nonumber\\
\lambda^{(c)}(Q) &=& \int du \cos{Qu }~ \lambda^{(c)}(u).
\end{eqnarray}
From here we can also straightforwardly conclude that we have the following identities for the functional
derivatives
\begin{eqnarray}
\frac{\delta g^{(c)}(Q)}{\delta g(s,s')} &=& 1 - \cos{Q (s - s')} \nonumber\\
\frac{\delta \lambda^{(c)}(Q)}{\delta \lambda(s,s')} &=& \cos{Q (s - s')}.
\end{eqnarray}
The next logical step in order to evaluate the partition function Eq. \ref{parti} would be to trace out the auxiliary
fields $\lambda(s), g_{ik}(s), {\cal B}_{ik}(s), p_{ik}(s), {\cal T}_{ik}(s)$. Unfortunately these fields
enter the
effective Hamiltonian non-linearly and they can not be simply traced over. Thus we have to introduce a new
approximation at this point:  instead of integrating over the auxiliary fields we will simply evaluate the
saddle - point of  $\overline{\cal H}'$ with respect to these fields. This is  easier and most importantly it
is feasible to do. It is thus at this juncture that a critical step in the
derivation of the partition function has to be made that constitutes the essence
of the $1/D$-expansion method as applied to the polymer
problem. The details and ramifications of this step have been addressed in the
previous work \cite{Hansen}.

\section{The saddle point equations}

The saddle point of the effective Hamiltonian Eq. \ref{eq.1.1} w.r.t. the
fields $ \lambda(s,s')$, ${\cal B}(s,s')$,  $g(s,s')$,  ${\cal T}(s,s') $ and 
$p(s,s')$ can be obtained straightforwardly in a standard fashion.
For the saddle point in the $ \lambda(s,s')$ variable we obtain the following relation
\begin{equation}
1 = \left( \partial_s {\bf r}_0(s) \cdot \partial_s {\bf r}_0(s)\right) + D k_B
T~ \sum_Q {\cal G}^ {-1}(Q) Q^2,
\label{eq.1.3}
\end{equation}
The next saddle point that we evaluate is ${\cal B}(s,s')$. It leads to the following equation
\begin{equation}
g^{(c)}(Q) = \int ds (1 - \cos{Qs}) \frac{\partial V({\cal B}(s), {\cal T}(s))}{\partial {\cal B}(s)}.
\end{equation}
Then follows the saddle point in $g(s,s')$, giving rise to
\begin{equation}
{\cal B}(s,s') =  \left( {\bf r}_0(s) -  {\bf r}_0(s')\right)^2 + 2D k_B T~
\sum_Q {\cal G}^{-1}(Q) (1 - \cos{Q (s - s')}).
\label{eq.1.4}
\end{equation}
Furthermore the saddle point in the variable ${\cal T}(s,s') $ leads to
\begin{equation}
p(Q) = \int ds \cos{Qs} \frac{\partial V({\cal B}(s), {\cal T}(s))}{\partial {\cal T}(s)},
\end{equation}
and finally the saddle point in $p(s,s')$ gives rise to
\begin{equation}
{\cal T}(s,s') = \left( \partial_s{\bf r}_0(s) \cdot \partial_s{\bf r}_0(s)\right) +
D k_B T~ \sum_Q {\cal G}^{-1}(Q) Q^2 \cos{Q (s - s')}.
\label{eq.1.5}
\end{equation}
We now assume that we can retain only terms up to the fourth order in $Q$ for the functions $\lambda^{(c)}(Q)$ and 
$g^{(c)}(Q)$. This is consistent with the fact that the original non-interaction part of the Hamiltonian contains only
Euler-Kirchhoffian elasticity and is thus at most of fourth order in the $Q$
space.  In this way we obtain the following expansion
\begin{equation}
\lambda^{(c)}(Q) = \lambda + p(Q)  = \lambda + \int ds  \frac{\partial V({\cal
B}(s), {\cal T}(s))}{
\partial {\cal T}(s)} - 
\frac{Q^2}{2}  \int ds s^2 \frac{ \partial V({\cal B}(s), {\cal T}(s))}{ \partial {\cal T}(s)} + \dots,
\end{equation}
as well as 
\begin{equation}
2 g^{(c)}(Q) = Q^2 \int ds s^2 \frac{\partial V({\cal B}(s), {\cal T}(s))}{\partial {\cal B}(s)} - 
\frac{Q^4}{12} \int ds s^4 \frac{ \partial V({\cal B}(s), {\cal T}(s))}{
\partial {\cal B}(s)} + \dots.
\end{equation}
Thus we have basically derived an expansion also for the self-energy Eq.
\ref{eq.1.2} which can now be cast into the form
\begin{equation}
{\cal G}(Q) = \left( \lambda + \delta \lambda \right) Q^2 + \left( K_C + \delta K_C \right) Q^4 +  \dots = 
\lambda^{(R)} Q^2 + K^{(R)}_C Q^4 +  \dots.
\end{equation}
The expansion in $Q $ to the fourth order thus allows us to introduce the renormalized values of the parameters
$ \lambda$ and $K_C $ via
\begin{equation}
\lambda^{(R)} = \lambda + \delta\lambda \qquad K^{(R)}_C = K_C + \delta K_C,
\end{equation}
where we introduced the following two abbreviations
\begin{eqnarray}
\delta \lambda &=& \int ds  \frac{\partial V({\cal B}(s), {\cal T}(s))}{\partial {\cal T}(s)} +
\int ds ~s^2 \frac{\partial V({\cal B}(s), {\cal T}(s))}{\partial {\cal B}(s)}
\nonumber\\
\delta K_C &=& - \ul12  \int ds~ s^2 \frac{\partial V({\cal B}(s), {\cal
T}(s))}{\partial {\cal T}(s)} - \textstyle{\frac{1}{12}} \int ds ~s^4
\frac{\partial V({\cal B}(s), {\cal T}(s))}{\partial {\cal B}(s)} .
\label{eq.1.7}
\end{eqnarray}
At this point we are in a position to evaluate all the $Q$ integrals in the saddle point equations. Thus instead of 
Eqs. \ref{eq.1.3}, \ref{eq.1.4} and \ref{eq.1.5} we remain with
\begin{eqnarray}
1 &=&  \left( \partial_s {\bf r}_0(s) \cdot \partial_s {\bf r}_0(s)\right) + 
\frac{D k_B T~}{4 \sqrt{\lambda^{(R)} K^{(R)}_C}}, \nonumber\\
{\cal B}(s,s') &=&  \left( {\bf r}_0(s) -  {\bf r}_0(s')\right)^2 + \frac{D k_B T~}{2 \lambda^{(R)}} \left(  (s-s') + \sqrt{\frac{K^{(R)}_C}{\lambda^{(R)}}} ( e^{-\sqrt{\frac{\lambda^{(R)}}{K^{(R)}_C}}(s- s')} - 1 )\right)  \nonumber\\
{\cal T}(s,s') &=& \left( \partial_s{\bf r}_0(s) \cdot \partial_s{\bf r}_0(s)\right)
 + \frac{D k_B T~}{4 \sqrt{\lambda^{(R)} K^{(R)}_C}} e^{-\sqrt{\frac{\lambda^{(R)}}{K^{(R)}_C}}(s- s')}.
\label{saddle}
\end{eqnarray}
Finally we have to address the question of the reference state. Assuming with
sufficient generality that the reference configuration of the chain is a
straight line, thus ${\bf r}_0(s) = \zeta s {\bf e}$, where ${\bf e}$ is a
constant vector \cite{Hansen}. At this point we can investigate also the saddle point of
$\overline{\cal H}[{\bf r}_0(s)]$ with respect to $\zeta$, the new variable,
which  reduces to the following   two equations
\begin{equation}
\frac{\partial{\cal G}(Q)}{\partial Q^2} = 0 \quad {\rm or} \quad \zeta = 0.
\end{equation}
On general grounds $\zeta$ can be non-zero only at zero temperature,
when there are no fluctuations and the chain can indeed exhibit straight
configurations. For any finite temperature straight configurations are not
feasible without external constraints and we should rather have $\zeta = 0$.
Taking this into account and solving the first of the above equations {\sl i.e.}
we obtain
\begin{equation}
1 =   \frac{d k_B T~}{4 \sqrt{\lambda^{(R)} K^{(R)}_C}}
\label{eq.1.6}
\end{equation}
Let us now  introduce a new parameter 
\begin{equation}
\xi = \frac{4 K^{(R)}_C}{D k_B T~}.
\end{equation}
In $D = 4$ $\xi$ would be simply equal the renormalized persistence length.
The factor $4/3$ obtained for $D = 3$ is a consequence of the $1/D$ ansatz,
{\sl i.e.} is a consequence of the approximate nature of the evaluation of the
partition function. We do not attribute much importance to this discrepancy.
With the introduction of $\xi$ equations Eqs. \ref{saddle} can be reduced to a
rather tame set of formulas
\begin{eqnarray}
{\cal B}(s) &=& 2 \xi \left( s + \xi (e^{-s/\xi} - 1) \right) \nonumber\\
{\cal T}(s) &=&  e^{-s/\xi}.
\label{eq.1.8}
\end{eqnarray}
These two formulas deserve some interpretation. The first of the above two
equations obviously represents the average size of the chain of contour length
$s-s'$ since at the saddle point ${\cal B}(s-s') = \mathopen< \left( {\bf r}(s)
- {\bf r}(s')\right)^2 \mathclose>$.  It is equal to the usual Kratky - Porod
expression for the average end-end distance 
squared of a semiflexible chain. The second equation simply expresses the fact,
that for a semiflexible chain the orientational correlation function, which at
the saddle point equals  $ {\cal T}(s-s') = \mathopen<  \partial_s{\bf r}(s)
\cdot    \partial_{s'}{\bf r}(s') \mathclose>$, is exponentially decaying with a
characteristic length equal to the persistence length. This again is completely
consistent with the Kratky - Porod result for a semiflexible chain
\cite{Rubinshtein}.

\section{Renormalized bending rigidity}

The above developments now allow us to analyze the explicit form of
the bending rigidity renormalization.  Taking first of all into account the definition of ${\cal
B}(s,s')$ and  ${\cal T}(s,s')$ we can write the dipolar interaction potential 
on the saddle point diagonalized level in the following form
\begin{equation}
V({\cal B}(s,s'), {\cal T}(s,s')) = V_R({\cal B}(s,s')) {\cal T}(s,s') - V_P({\cal B}(s,s')) {\cal B}(s,s') 
{\cal T}(s,s'),
\end{equation}
allowing us to evaluate $\delta \delta K_C$ as a function of the parameters of
this interaction potential. We  obtain
\begin{eqnarray}
\delta K_C &=& - \frac{1}{2} \int ds ~s^2~\left( V_R({\cal B}(s)) -  V_P({\cal B}(s)) {\cal B}(s)\right) - 
\nonumber\\
& & - {\frac{1}{12}} \int ds ~s^4~\left( \left( \frac{\partial V_R({\cal B}(s))}{\partial {\cal B}(s)} - 
V_P({\cal B}(s))\right){\cal T}(s)  - \frac{\partial V_P({\cal B}(s))}{\partial {\cal B}(s)}{\cal B}(s){\cal T}(s) 
\right).
\label{eq.1.7}
\end{eqnarray}
It is instructive to compare this with the analogous formulation of the problem for
monopolar interactions \cite{Hansen}. There it was derived that 
\begin{equation}
\delta K_C = -{\frac{1}{12}} \int ds ~s^4~ \frac{\partial V_R({\cal B}(s))}{\partial {\cal B}(s)}.
\end{equation}
In view of this, it seems that a possible interpretation of Eq. \ref{eq.1.7} would be that the orientational
part of the interaction potential averaged over chain conformational fluctuations
leads to {\sl effective} intersegment attractions with a repulsive as well as
attractive components.

An exact evaluation of the highly non-linear Eq. \ref{eq.1.7}, note that $\delta
K_C$ is on the l.h.s. of this equation as well as on the r.h.s., hidden in $\xi$
of the definitions Eq. \ref{eq.1.8}, is not feasible and we have to investigate
the properties of its solution numerically. First of all we break the evaluation
of the integral Eq. \ref{eq.1.7} into two parts $\delta K_C = \delta K_C^{(1)} +
\delta K_C^{(2)}$, defined as
\begin{eqnarray}
\delta K_C^{(1)} &=& -\frac{1}{2} \int_a^{\infty} ds~s^2~\left( V_R(\kappa\sqrt{{\cal B}(s)}) -  
V_P(\kappa\sqrt{{\cal B}(s)}) {\cal B}(s) \right) \nonumber\\
\delta K_C^{(2)} &=& - {\frac{1}{12}}  \int_a^{\infty} ds~s^4~\left(  \frac{\partial 
V_R(\kappa\sqrt{{\cal B}(s)})}{\partial {\cal B}(s)} - V_P(\kappa\sqrt{{\cal
B}(s)})  - \frac{
\partial V_P(\kappa\sqrt{{\cal B}(s)})}{
\partial {\cal B}(s)}{\cal B}(s) \right) {\cal T}(s), \nonumber\\
~
\end{eqnarray}
where we have taken into account that the interaction potential can be written
as a function of the argument $\kappa  \vert{\bf r}(s) - {\bf r}(s')\vert$.
Furthermore the lower cutoff was set equal to $a$, of the order of the thickness
of the chain, whose numerical value was taken as $a = 1 nm$. This cutoff stemms from the breakdown of the continuum elasticity at small
lengthscales. Contrary to the monopolar case \cite{Hansen}, this cutoff is 
essential and reflects the faster
decay of the dipolar interactions with respect to the separation compared to
the monopolar case. The next step is to write above relations in the form
suitable for numerical evaluation.
\begin{figure}[ht]
\begin{center}
    \epsfig{file=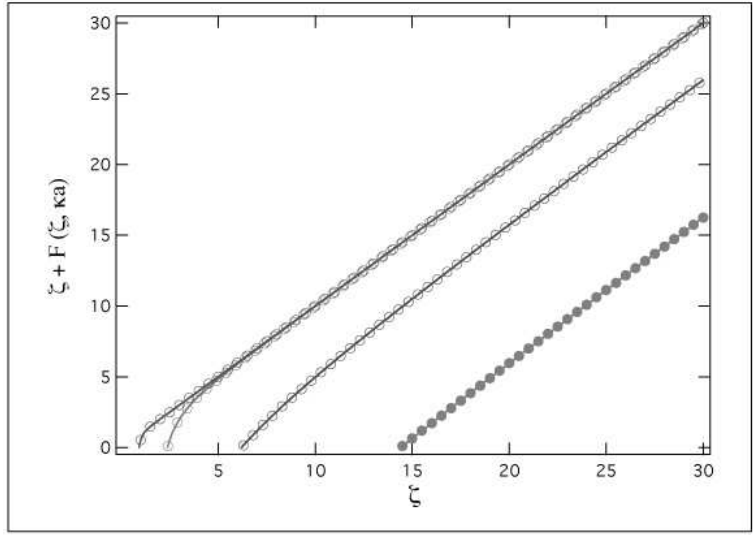, width=10 cm}
\end{center}    
\caption{The $\zeta + F(\zeta)$ Eq. \ref{def-f} for four different values of
$\kappa a$, {\sl viz.} $\kappa a = 100, 10, 1$ and $0.1$ (from left to right). The dimensionless strength of the dipolar
interaction $\left(\kappa \ell_B\right) \left( \frac{p_0}{e_0} \right)^2 $ was
taken as $1$. At large values of $\zeta$ the dependence is obviously linear.} 
\label{fig0}
\end{figure}
Let us first of all introduce the following notation for the interaction potential
\begin{eqnarray}
V_R(r) &=& \frac{p_0^2}{4\pi \epsilon \epsilon_0} \frac{e^{-\kappa r}}{r^3} (1 + \kappa r) = v_0 \kappa^3 f_R(\kappa r) \nonumber\\
V_P(r) &=& \frac{p_0^2}{4\pi \epsilon \epsilon_0} \frac{e^{-\kappa r}}{r^5} (3 + 3\kappa r + (\kappa r)^2)) = v_0 \kappa^5 f_P(\kappa r),
\end{eqnarray}
where $f_{R,P}(u)$ are obviously defined as
\begin{eqnarray}
f_R(u) = \frac{e^{-u}}{u^3} (1 + u) \nonumber\\
f_P(u) = \frac{e^{-u}}{u^5} (3 + 3u + u^2). 
\end{eqnarray}
Introducing furthermore
\begin{equation}
\overline{\cal B}(z) = z - 1 + e^{-z} \quad {\rm with} \quad z = \frac{s}{\xi},
\end{equation}
and setting 
\begin{equation}
\overline{\xi} = 1/(\sqrt{2} \kappa \xi), 
\end{equation}
we can derive the following form for $\delta K_C^{(1)}$
\begin{equation}
\delta K_C^{(1)} = - {\frac{v_0}{2}} (\kappa \xi)^3 \int_{a/\xi}^{\infty} dz z^2 \left( 
f_R \left( \sqrt{\overline{\cal B}(z)}/\overline{\xi} \right) - f_P\left( 
\sqrt{\overline{\cal B}(z)}/\overline{\xi} \right) \frac{\overline{\cal B}(z)}{\overline{\xi}^2} 
\right).
\label{eq.2.0}
\end{equation}
and for $\delta K_C^{(2)}$
\begin{equation}
\delta K_C^{(2)} = - \frac{v_0}{12} (\kappa\xi)^5  \int_{a/\xi}^{\infty} dz z^4 \left( f'_R\left( 
\sqrt{\overline{\cal B}(z)}/\overline{\xi}\right) - f_P\left( \sqrt{\overline{\cal 
B}(z)}/\overline{\xi} \right) - f'_P\left( \sqrt{\overline{\cal B}(z)}/\overline{\xi} \right) 
\frac{\overline{\cal B}(z)}{\overline{\xi}^2}\right) ~e^{-z},
\label{eq.2.1}
\end{equation}
where the derivative in $f'_R$ and $f'_P$ stands for $f' = \frac{\partial f(\sqrt{u})}{\partial u}$ and thus we have 
\begin{eqnarray}
f'_R(u) &=& -\frac{e^{-u}}{2 u^5 } (3 + 3u + u^2) \nonumber\\
f'_P(u) &=& -\frac{e^{-u}}{2 u^7 } (15 + 15u + 6u^2 + u^3).
\end{eqnarray} 
We should again note here that the equations for the renormalized bending modulus
Eqs. \ref{eq.2.0} and \ref{eq.2.1} are  essentially non-linear, since $\delta
K_C$ is also hidden in the variable $\overline{\xi} = d k_B T~ /(4 \sqrt{2} 
\kappa (K_C + \delta K_C))$. 
\begin{figure}[ht]
\begin{center}
    \epsfig{file=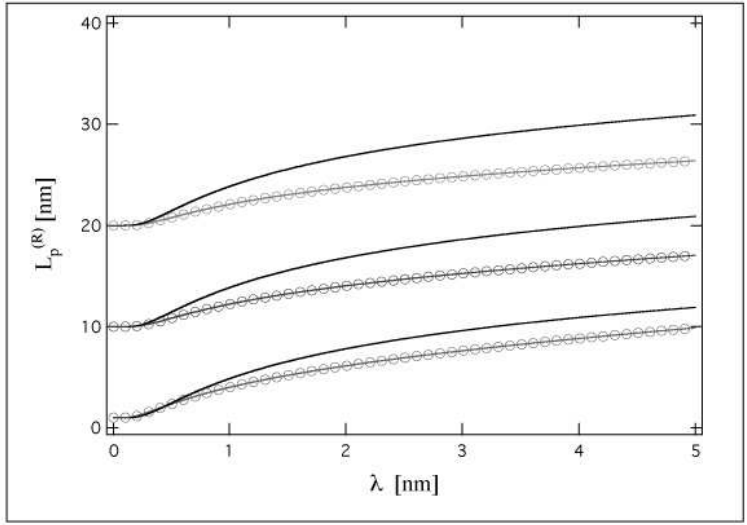, width=10 cm}
\end{center}    
\caption{Renormalized persistence length Eq. \ref{finaleq} for three different values of $L_p^{(0)} =
20, 10$  and $1$nm (upper, middle, lower curves), circles. Also shown are the
approximate solutions obtained from the expansion Eq. \ref{approx-2}.  The
strength of the dipolar interaction $\ell_B \left(
\frac{p_0}{e_0} \right)^2 $ was taken as $1$ nm.}
\label{fig1}
\end{figure}
Let us finally introduce a new variable $\zeta$ defined as
\begin{equation}
\zeta = \overline{\xi}^{-1} = \sqrt{2} \kappa \xi = \frac{(4 \sqrt{2} 
\kappa K_C^{(R)})}{3 k_B T~} = \frac{4 \sqrt{2}}{3} \frac{L_p^{(R)}}{\lambda_{D}},
\label{defzeta}
\end{equation}
where $\lambda_{D}  = 1/\kappa$ is the Debye screening length. $\zeta$ 
is nothing but the inverse reduced screeing length as introduced by Everaers 
{\sl et al.} \cite{Everaers}. Furthermore if
\begin{equation}
\zeta^{(0)} = \frac{4 \sqrt{2} \kappa K_C}{d k_B T~} \qquad {\rm and} \qquad \delta\zeta = 
\frac{4 \sqrt{2}  \kappa ~\delta K_C}{d k_B T~}. 
\end{equation}
then the definition of the renormalized bending rigidity $K_C^{(R)} = K_C +
\delta K_C$ can be cast into the form
\begin{equation}
\zeta = \zeta^{(0)} - \frac{\kappa v_0}{D k_B T~} \zeta^3 \left( F^{(1)}(\zeta) + \frac{\zeta^2}{12}
F^{(2)}(\zeta) \right).
\end{equation}
Because of Eq. \ref{defv0} we can furthermore write in 3 dimensions
\begin{equation}
\zeta + F(\zeta) = \zeta + \ul13 \left(\kappa \ell_B\right) \left( \frac{p_0}{e_0} \right)^2 
\zeta^3 \left(F^{(1)}(\zeta) + {\textstyle{\frac{1}{12}}} \zeta^2 F^{(2)}(\zeta) \right) =
\zeta^{(0)}.
\label{def-f}
\end{equation}
This  equation gives the functional dependence of $\zeta$ on the parameters 
of the dipolar interaction, most notably $\lambda_{D} $. From the definition 
of $\zeta$ Eq. \ref{defzeta} the solution of the above equation immediately
leads to the renormalized value of the persistence length. The functions
$F^{(i)}(\zeta)$ have obviously been defined as
\begin{eqnarray}
F^{(1)}(\zeta) &=& \int_{\sqrt{2} (\kappa a)/\zeta}^{\infty} dz~ z^2 \left( 
f_R \left( \zeta\sqrt{\overline{\cal B}(z)}\right) - f_P\left( 
\zeta \sqrt{\overline{\cal B}(z)}\right) \zeta^2 \overline{\cal B}(z) 
\right) \nonumber\\
F^{(2)}(\zeta) &=&  \int_{\sqrt{2} (\kappa a)/\zeta}^{\infty} dz ~z^4 \left( f'_R\left( 
\zeta \sqrt{\overline{\cal B}(z)}\right) - f_P\left( \zeta\sqrt{\overline{\cal 
B}(z)}\right) - f'_P\left( \zeta\sqrt{\overline{\cal B}(z)}\right) 
\zeta^2 \overline{\cal B}(z)\right) ~e^{-z}. \nonumber\\
~
\label{integrals}
\end{eqnarray}
Fig. \ref{fig0} shows the dependence of $\zeta + F(\zeta)$ on $\zeta$ for four
different values of $(\kappa a)$ that enters the definition of $F(\zeta)$ {\sl
via} the lower bound of the integrals  Eq. \ref{integrals}. The solution of
Eq. \ref{def-f} is obtained from Fig. \ref{fig0} by simply looking at the
intersection between the curves on the figure with a line parallel to the
$\zeta$ axis at $\zeta^{(0)}$.
\begin{figure}[ht]
\begin{center}
    \epsfig{file=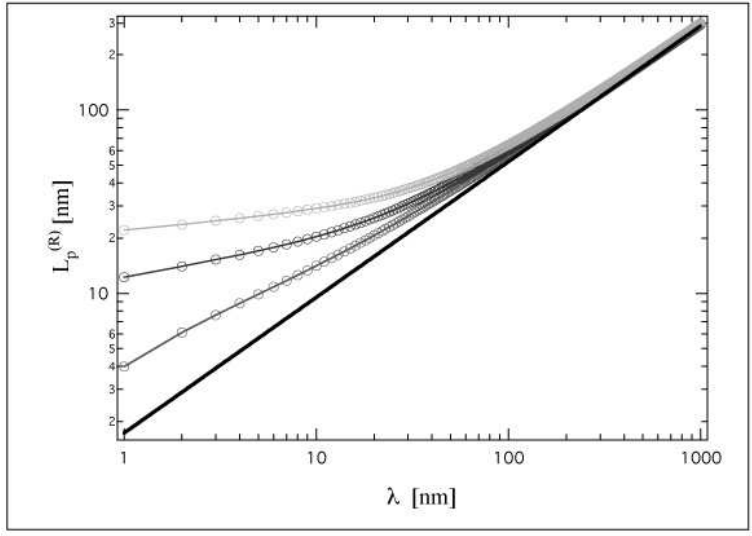, width=10 cm}
\end{center}    
\caption{Fit (bold line)) of the renormalized persistence length to a power of
$\lambda^\alpha$ for three
different values of $L_p^{(0)} = 1, 10$ and $20$ nm (lower, middle and upper circles).
The scaling exponent obtained from the fit is $\alpha = 0.75$. The
strength of the dipolar interaction $\ell_B \left(
\frac{p_0}{e_0} \right)^2 $ was taken as $1$ nm.}
\label{fig2}
\end{figure}
A more transparent form of these equations that will be solved numerically can be
obtained if we go back to the original variables and can thus write for the
renormalized persistence length 
\begin{equation}
L_p^{(R)} = L_p^{(0)} - \frac{\ell_B}{4 \sqrt{2}} \left(\frac{p_o}{e_0}\right)^2 \left( 
\frac{4 \sqrt{2}}{3} \frac{L_p^{(R)}}{\lambda_{D}}\right)^3 \left(
F^{(1)}\left(\frac{4 \sqrt{2}}
{3} \frac{L_p^{(R)}}{\lambda_{D}}\right) + {\textstyle{\frac{1}{12}}} \left(
\frac{4 \sqrt{2}}{3} 
\frac{L_p^{(R)}}{\lambda_{D}}\right)^2 F^{(2)}\left(\frac{4 \sqrt{2}}{3}
\frac{L_p^{(R)}}{\lambda_{D}}
\right) \right).
\label{finaleq}
\end{equation}
This equation is the main result of our paper. Its solution gives the
renormalized value of the bending rigidity or equivalently the persistence
length as a function of the dipolar interaction parameters. In general it can
only be solved numerically but simplified analytic solutions can be obtained in
limited regions of the paramater phase space.

It is instructive to compare these results with those derived for a
polyelectrolyte chain  with simple screened  Debye - H\" uckel monopolar
electrostatic interactions along the chain, where the fixed charges are at
separation $A$ along the chain. If we formulate the
results derived in in the same language as used above, we get 
\begin{equation}
\zeta = \zeta^{(0)} + \frac{\sqrt{2} \kappa\ell_{B}}{12 ~2^{3} (A\kappa)^{2} }
\zeta^{2} F(\zeta), 
\end{equation}
where in this case
\begin{equation}
F(\zeta) = \int_{0}^{\infty} dz z^{4} \left( 1 + \zeta \sqrt{\overline{\cal
B}(z)}\right) \frac{e^{-\zeta }\sqrt{\overline{\cal B}(z)}}{{\overline{\cal
B}(z)}^{{3/2}}}.
\end{equation}
The analysis of these equations (though written in a somewhat different, yet completely equivalent form) has
already been performed before and we direct the reader to that work \cite{Hansen}.

\section{Results}

We noted already that the basic equations derived for the renormalization of the
persistence length of a semiflexible  dipolar chain have no simple analytic
solution. Figures \ref{fig1} and \ref{fig2} thus present numerical 
solutions of Eq. \ref{finaleq} for some values of parameters  $L_p^{(0)}$,
$\lambda_{D} $, while we always consider the case  $\frac{\ell_B}{2}
\left(\frac{p_o}{e_0}\right)^2 = 1 nm^{-1}$. This last parameter sets the
overall scale of the renormalization and is not crucially significant so we do
not study its variation in detail. Two universal behaviors seem to emerge. For
small values of the screenenig length $\lambda_{D}$ the  behavior is given in
Fig. \ref{fig1}. For large values of the screening length an altogether different
type of behavior is seen Fig. \ref{fig2}. Numerical solutions of  Eq. \ref{finaleq}
do not show any indications of a possible phase transition to an ordered phase
at non-zero temperatures though due to the attractive component of the dipolar
interactions one might expect them to. This is completely consistent with the
behavior of the flexible chain \cite{Muthu}.  

Let us investigate if we can understand the two types of behavior depicted on
Figs. \ref{fig1} and \ref{fig2} with simple analytical arguments.
What we basically need to figure out is what would be the typical contributions
to the integrals Eq. \ref{integrals}  that enter the equation for the
renormalized bending rigidity Eq. \ref{finaleq}. It is instructive to consider
first the case of small screening lengths. Here the main contribution to the
integrals Eq. \ref{integrals} along the polyelectrolyte comes from short length
scales.  In this limit one derives that  ${\cal B}(s) \simeq {s^2}$ and locally
the chain thus behaves in the Kratky - Porod style \cite{Rubinshtein}. This
leads to the following approximate form of $F(\zeta)$
\begin{equation}
F(\zeta) = \zeta^3\left( F^{(1)}(\zeta) + \frac{\zeta^2}{12 }F^{(2)}(\zeta) \right) = 2 \sqrt{2} 
\int_{\sqrt{2}(\kappa a)/\zeta}^{\sqrt{2}/\zeta} \frac{dz}{z} \left( -18 - 18 \frac{\zeta z}{\sqrt{2}} - 9
\left(\frac{\zeta z}{\sqrt{2}}\right)^2 + \left(\frac{\zeta z}{\sqrt{2}}\right)^3 \right).
\label{approx-1}
\end{equation}
Eq. \ref{finaleq} in this very same limit then assumes the form
\begin{equation}
L_p^{(R)} = L_p^{(0)} - \frac{\ell_B}{2} \left(\frac{p_o}{e_0}\right)^2   \left(
\frac{(-25 -7(\kappa a) + 
(\kappa a)^2) e^{-(\kappa a)}}{12} - \frac{3}{2} E_1(\kappa a) \right),
\label{approx-2}
\end{equation}
where $E_1(u)$ is the standard exponential integral with a limiting form of
$\lim_{u \longrightarrow 0}E_1(u) = -  \gamma - \log{u} + {\cal O}(u)$. 
Except for very small values of the screening length where obviously $L_p^{(R)}
\approx L_p^{(0)}$, a valid approximation of Eq. \ref{approx-2} capturing its 
dominant features would be
\begin{equation}
L_p^{(R)} \approx L_p^{(0)} + \frac{3 \ell_B}{4} \left(\frac{p_o}{e_0}\right)^2  
\log{\frac{\lambda_{D}}{a}}.
\label{final-1}
\end{equation}
This result could be seen as the dipolar chain equivalent of the Odijk - Skolnick
- Fixman \cite{OSF} result for the monopolar chain. In that case the behavior of
the persistence length at small values of the screening length is $L_p^{(R)}
\sim \lambda_{D}^{2}$. The difference between this result and Eq. \ref{final-1}
is purely due to the swifter decay of the dipolar interaction potential vs.
separation, $r$, in monopolar case $r^{-1}$ and in dipolar case $r^{-3}$.  
This can be seen crudely as follows: since the interaction potential for the
dipolar chain falls off twice as fast as the monopolar potential the OSF
reasoning should give for the electrostatic persistence length $
\lambda_{D}^{2} \times  \lambda_{D}^{-2} =  \lambda_{D}^{0}$. The zero in the
exponent translates as usual into the $log$, which is exactly Eq.
\ref{final-1}. The way the approximate result Eq. \ref{approx-2}  fares when 
compared with the full numerical solution of Eq. \ref{finaleq} is shown on Fig.
\ref{fig1}. Obviously the approximaton is not bad for small enough screening
length but becomes progressively worse for larger bare persistence lengths and
larger screening. The same reasoning applied to the general multipole of order
$n$ would lead to the electrostatic persistence length with scaling
$\lambda_{D}^{-(n-2)}$. One can thus hardly expect any electrostatic
renormalisation effects above dipolar.

Next we consider the case of large(er) screening lengths. Here the main
contribution to the integrals Eq. \ref{integrals} along the polyelectrolyte
comes from large(er) length scales.  In this case the local properties of the
chain are characterized by an essentially free flight behavior
\cite{Rubinshtein} of the chain, leading to  ${\cal B}(s) \simeq 2 \xi s$. Here
we can derive 
\begin{equation}
F(\zeta) = \zeta^3\left( F^{(1)}(\zeta) + \frac{\zeta^2}{12 }F^{(2)}(\zeta)
\right) = - \int_{\sqrt{2} (\kappa a)/\zeta}^{1/\zeta^2} dz z^{1/2} \left( 2 + 2
\zeta\sqrt{z} + \zeta^2 z\right) = - \frac{41}{15} 
\zeta^{-3},
\label{approx-3}
\end{equation}
which leads to the following approximate form of Eq. \ref{finaleq} 
\begin{equation}
L_p^{(R)} \approx L_p^{(0)} + \frac{\ell_B}{4 \sqrt{2}}
\left(\frac{p_o}{e_0}\right)^2
 \frac{41}{15} 
\left(\frac{4 \sqrt{2}}{3} \frac{L_p^{(R)}}{\lambda_{D}}\right)^{-3} = L_p^{(0)}
+ 0.07 ~\ell_B \left(\frac{p_o}{e_0}\right)^2
\left(\frac{\lambda_{D}}{L_p^{(R)}}\right)^3.
\label{approx-4}
\end{equation}
This equation can have two possible solutions depending on the magnitude of
$L_p^{(0)}$. First of all if $L_p^{(R)} \gg L_p^{(0)}$, meaning that the value
of the persistence length is determined mostly by the electrostatic interactions
along the chain, we have
\begin{equation}
L_p^{(R)} = 0.07^{1/4}\left( \ell_B \left(\frac{p_o}{e_0}\right)^2\right)^{1/4}
\lambda_{D}^{3/4}.
\label{limit-1}
\end{equation}
This approximate result is nicely exhibited also on Fig. \ref{fig2} where it
is seen how the exact numerical solutions of Eq. \ref{finaleq} approach the
above scaling limit for sufficiently large values of the screening length with
the scaling exponent exactly equal to the above prediction. The form Eq.
\ref{limit-1} is thus a limiting law for the electrostatic part of the
persistence length valid universally for large screening lengths. One should
be however aware, that in the above calculation by assumption the length of the
chain is always the largest length in the system.

In the other case with a predominant contribution of the bare persistence
length, {\sl i.e.} $L_p^{(R)} \sim L_p^{(0)}$, thus in the limit where the
strength of the dipolar interaction is very small, we end up with the
following limiting form
\begin{equation}
L_p^{(R)} = L_p^{(0)} + 0.07~ \ell_B
\left(\frac{p_o}{e_0}\right)^2\left(\frac{\lambda_{D}}{L_p^{(0)}}\right)^3.
\label{limit-2}
\end{equation}
The latter  is only valid for small enough strength of the dipolar interaction
and is difficult to discern in the numerical solution. Since all our numerical
investigations on Figs. \ref{fig1}, \ref{fig2} are based on the assumption of a
fairly large strength of the dipolar interactions, this regime is not
distinguishable on the graphs. 

These last two results could again be compared with  the expressions valid for
the monopolar chain in the same regions of the parameter space and derived within
the same theoretical framework \cite{Hansen}. In that case the behavior of the persistence
length at large values of the screening length would be $L_p^{(R)} \sim
\lambda_{D}^{\beta}$, where $\beta $ is either $7/6$ or $7$, depending on whether
$L_p^{(R)}/L_p^{(0)}$ is either large or small. Again the difference between
these results and Eqs. \ref{limit-1} and \ref{limit-2} is purely due to the
swifter decay of the dipolar interaction potential vs. separation; monopolar case
$r^{-1}$ and dipolar case $r^{-3}$.  This difference introduces an additional
factor of $\lambda_{D}^{-2} ~ \left(\frac{\lambda_{D}}{L_p^{(R)}} \right)^{-2}$
into the second term of Eq. \ref{approx-4}, wherefrom the exponents $\beta = 3/4$
or $3$ can be derived straightforwardly. It is thus clear that the behavior of
the electrostatic part of the persistence length of a monopolar and dipolar
chain are intimately related in both relevant limits of weak and strong
screening. Applying again the same reasoning to the general multipole
of order $n$ would lead to the electrostatic persistence length with scaling
$\beta =  (7-2n)$ or $\beta = (7-2n)/(6-n)$. Again electrostatic
renormalisation effects above dipolar are very small.

The remaining question in this context would be how robust are the two regimes
derived above that lead to the approximate forms Eqs. \ref{final-1} and
\ref{limit-1}? In assessing the range of validity of these different
approximations we can invoke a related situation already encountered in the case
of a monopolar chain \cite{Hansen}. In that system extensive simulations
\cite{Everaers, Ullner, Shklovski} left no doubt that
the OSF regime, corresponding in the context of the dipolar chain to Eq.
\ref{final-1}, is very robust and extends over a broad region of the parameter
phase space. The sub-OSF laws for the electrostatic renormalization of the
persistence length giving $\beta$ in the vicinity of $1$ were effectively ruled
out. Translating this into the context of the dipolar chain would make the
range of validity fo Eq. \ref{approx-4} fairly narrow. But these are all
conjectures since we are aware of no detailed simulations for semiflexible
dipolar chains, setting aside the extensive work by Muthukumar on the {\sl
flexible} chain \cite{Muthu}, that would match the superb work perfomed recently
in the context of the monopolar chain \cite{Everaers, Ullner, Shklovski}. 

\section{Conclusions}

We presented an analysis of the electrostatic contribution to the persistence
length of a semiflexible screened-dipolar chain. The formal context of our
analysis is provided by the $1/D$-expansion method that has been already
successfully applied to different problems of polymer physics \cite{Hansen,
Podgornik, Hansen2}. $1/D$-expansion
is closely related to different variational schemes that have been
amply applied to the problem of electrostatic rigidity of charged polymers
\cite{Podgornik2, Bratko, LiWitten, Netz, baeyun}. What singles out our approach
is the
fact that we work consistently with a semiflexible Hamiltonian and enforce the
inextensibility constraint on a global level. This leads to some discrepancies
between variational formulations and $1/D$-expansion method. However the
robustness of the OSF  regime transpires quite clearly from the $1/D$-expansion
in the context of a monopolar polyelectrolyte, giving us some confidence that an
analogous result for a dipolar chain would have the same range of validity. 

The main step in our formulism was to find an appropriate way to treat the
orientational dependence of the intersegment interaction potential. This has
been accomplished by introducing additional auxiliary fields that in their
turn lead to new saddle-point equations. Though the derivation of the
renormalized elasticity is as a consequence more complicated, it leads to
manageable and transparent results. The main conclusion based on these
results would be that the dependence of the electrostatic persistence
length on the screening length for a dipolar chain is much less pronounced then
for a monopolar chain. This is clearly seen in the case of large as well as
small screening. Also it appears that a semiflexible chain does not give rise
to any localized structures as those described by Muthukumar  \cite{Muthu} in the
context of a flexible chain. This is indeed not surprising since the chain
elasticity is obviously strong enough to prevent extensive looping of the chain
that would lead to local aggregation of the chain based on the attractive
component of the dipolar interaction. This attractive component is also not
strong enough to lead to any type of phase transition to an ordered state at non-zero
temperatures. This conclusion is just as valid for a semiflexible as it
is for a flexible chain. 

Eventually the conclusions arrived at in this work would have to be tested
against extensive simulations just as they were in the case of a monopolar
charged chain. As for the experimental situation where repeated claims of a sub-OSF regime
have been voiced  \cite{Reed, Muthu2}, one wonders if they could in fact be a consequence of a
specific association of the mobile counterions with fixed charges on the
polyelectrolyte backbone producing a system not far from the model dipolar
chain analyzed in this work. In order to test this hypothesis one would have
to estimate independently the effective dipole moments of the chain segments. The
results presented in this work could then serve as a guideline to differentiate
between monopolar and dipolar OSF-like behavior. 

\vskip 1cm

{\large\bf Acknowledgment} I would like to thank Per Lyngs Hansen for numerous
discussions regarding polyelectrolytes and their theoretical description.

\end{document}